\documentclass[]{elsart}

\usepackage{graphicx}
\usepackage{isolatin1}

\usepackage{amsmath}
\usepackage{amssymb}
\usepackage{bm}
\journal{Journal of Computational Physics}
\begin{document}

\begin{frontmatter}
\title{Variational multiscale turbulence modelling in a high order 
       spectral element method}%
\author{Thor Gjesdal\corauthref{cor}},
 \corauth[cor]{Corresponding author.}\ead{Thor.Gjesdal@ffi.no} %
\author{Carl Erik Wasberg}\ead{Carl-Erik.Wasberg@ffi.no},  %
\author{Bj\o rn Anders Pettersson Reif}\ead{Bjorn.Reif@ffi.no}, %
\author{\O yvind Andreassen}\ead{Oyvind.Andreassen@ffi.no}
\address{Norwegian Defence Research Establishment (FFI), P.O. Box 25,
 NO-2027 Kjeller, Norway}
\date{25 May 2006}

\begin{abstract}
One of the more promising recent approaches to turbulence modelling is
the Variational Multiscale Large Eddy Simulation (VMS LES) method
proposed by Hughes et al. [Comp. Visual. Sci., vol. 3, pp. 47-59,
2000]. This method avoids several conceptual issues of traditional
filter-based LES by employing a priori scale partitioning in the
discretization of the Navier-Stokes equations.

Most applications of VMS LES reported to date have been based on
hierarchical bases, in particular global spectral methods, in which
scale partitioning is straightforward. In the present work we describe
the implementation of the methodology in a three-dimensional
high-order spectral element method with a nodal basis. We report
results from coarse grid simulations of turbulent channel flow at
different Reynolds numbers to assess the performance of the model.
\end{abstract}

\begin{keyword}
Large eddy simulation \sep variational multiscale method \sep 
spectral element method \sep incompressible flow
\MSC 76F65 \sep  65M60 \sep 65M70
\PACS 47.27.Eq \sep 02.70.Hm \sep 02.70.Dh
%The PACS codes are: 
% + 47.27.Eq: Turbulence modelling and simulation
% + 02.70.Hm: Spectral methods
% + 02.70.Dh: Finite element and Galerkin methods
\end{keyword}
\end{frontmatter}

% main text
\section{Introduction}

Large-eddy simulations (LES) provides a physically  more appealing
framework for turbulent flow prediction than the more traditional
Reynolds-averaged models (RANS). In the latter the full impact of the
ensemble averaged effect of turbulent advection on the mean flow field
has to be modeled. The essence of the LES approach on the other hand is 
to directly solve (with a complete time and space resolution) the
three-dimensional and time-dependent motion of the largest turbulent
scales. These scales are in general associated with the most energetic
motion of the turbulence field and it is (ideally) only the least
energetic motion that need to be modeled. The concept as such is
therefore well suited to confront the scale complexity and transient
behavior inherent to turbulent flows and offers a more complete
representation than RANS models per se.  

In traditional LES, large- and small-scale motion are separated by
applying a spatial filtering operation to the Navier-Stokes equations.
This results in a set of equations for the large-scale motion. The
residual motion, i.e.\ turbulent motions on scales that are smaller
than the filter width, appear in these equations as a residual stress
term that must be modeled. There are several conceptual issues in
filter-based LES that have to be addressed. For instance, filtering
and spatial differentiation do not in general commute on bounded
domains or for non-uniform grids, and it is not obvious how to
prescribe correct boundary conditions for the filtered velocity at
solid walls.  Another unwarranted character of filter-based LES models
is that the residual stress model has a tendency to affect the
entire range of the spectrum and not only represent the filtered
effect of the unresolved scales near the spectral cut-off.  These
issues have been the subject of a considerable amount of research, and
the lesson learned, in general, is that LES works well in cases where
the rate-controlling processes occur at the largest (resolved) scales
of motion, or equivalently in flows where the unresolved scales, and
consequently the model, only plays a secondary dynamical role.

In this paper we consider a different approach to LES, the variational
multiscale (VMS) LES method originally proposed by Hughes
\etal~\cite{TJRHughes_LMazzei_KEJansen_2000a}.  
The VMS LES method employs an a priori scale partitioning in the
discretization of the Navier-Stokes equations, instead of filtering to
separate the large- and small-scale motion.  
The scale partitioning appears to overcome some of the
disadvantages of filter-based LES. First, since there is no filtering,
all issues concerning commutation errors and boundary conditions at
solid walls are addressed. Second, since the scale partitioning is
performed during discretization, we develop different equations
representing different ranges of the spectrum. Different modelling
assumptions can then be applied to each range of the spectrum,
improving our ability to apply the model terms where they are needed,
and only there.

We implement the VMS LES formulation in a high order spectral element
method for the solution of the Navier-Stokes equations. Spectral
element methods offer an attractive combination of the accuracy of
spectral methods and the flexibility of finite element methods. This
provides us with an attractive framework for model development in
which the numerical errors can be controlled, such that the true
performance of the model can be assessed.  The first implementations
of the variational multiscale LES method~%
\cite{TJRHughes_LMazzei_AAOberai_AAWray_2001a,TJRHughes_AAOberai_LMazzei_2001a,SRamakrishnan_SSCollis_2002b}
used global spectral methods. These methods naturally employ an
orthogonal modal basis, such that the scale partitioning becomes
straightforward. Recently, the method has also been implemented in the
context of other numerical schemes, such as finite element methods~%
\cite{VGravemeier_etal_2004a,VJohn_SKaya_2005a} 
and finite volume methods~%
\cite{CFarhat_etal_2003a}.  
Our spectral element code uses an element-wise discretization with
nodal basis functions that contain information on all the scales. One
of the challenges of the present work is therefore to devise a way to
separate the large and the small scales, and to implement the VMS
terms. We show that this can be achieved by an element-by-element
transformation into the Legendre modal basis functions.

In the following sections we will discuss the variational multiscale
method as a turbulence modelling tool, and describe the implementation
of the method in a spectral element solver for the incompressible
Navier-Stokes equations. Finally we will present computed results,
from both a high-resolution DNS and coarse grid VMS LES for the
turbulent flow in a plane channel at different Reynolds numbers. The
computed results show that, even with simple modelling applied to the
small-scale equations, the performance of the methodology is
promising.

% -----------------------------------------------------------------------------
\section{The variational multiscale method}
\label{sec:vms}
In this section we will discuss the variational multiscale method as a tool for
turbulence modelling. The variational multiscale LES method was
introduced by Hughes \etal~\cite{TJRHughes_LMazzei_KEJansen_2000a}  
and later elaborated by Collis~\cite{SSCollis_2001a}. We will outline
the method following Collis, to shed light on the modelling
assumptions employed in the derivation of the model.

The Navier-Stokes equations describing the dynamics of a viscous, 
incompressible fluid are
\begin{subequations}
\label{eq:navier-stokes}
\begin{gather}
\nabla\cdot\bm{u} = 0, \\
\frac{\partial \bm{u}}{\partial t} + \bm{u}\cdot\nabla{\bm{u}} = 
   -\nabla{p} + \nu\nabla^2 \bm{u} + \bm{f}, 
\end{gather}
\end{subequations}
where the independent variables are the velocity, $\bm{u}=(u,v,w)$, and the
pressure, $p$.
The kinematic viscosity is denoted by $\nu$, and $\bm{f}$ is a body force
term.  The non-dimensional parameter
that characterizes the flow is the Reynolds number $\mathrm{Re}=|\bm{u}|L/\nu$.

For ease of presentation we assume homogeneous Dirichlet boundary
conditions for the velocity, i.e.
\[ \bm{u}(x) = 0 \quad x\in\Gamma. \]
We can then construct the weak, or variational, formulation by
choosing test and trial functions in the same function space
$\mathcal{V}$. Note however that in general the test and trial spaces
will differ at the boundary.
\[
\begin{aligned}
U &= (\bm{u},p) &\in \mathcal{V} \\
W &= (\bm{w},q) &\in \mathcal{V} 
\end{aligned}
\]
We take the inner product of $W$ with Eq.~%
\eqref{eq:navier-stokes} 
%over the space-time slab $Q=\Omega\times]0,T[$ 
(written in the compact form $\mathcal{N}(U) = F$) to obtain the weak
Navier-Stokes operator: 
\begin{equation}
  (W, \mathcal{N}(U)) \equiv \mathcal{L}(W,U) -
  \mathcal{R}(\bm{w},\bm{u}) = (W,F), 
\end{equation}
comprising the linear Stokes operator
\begin{equation}
  \mathcal{L}(W,U) \equiv (\bm{w},\frac{\partial \bm{u}}{\partial t})
  - (\nabla \cdot \bm{w}, p) + 
  (\nabla^s \bm{w}, 2 \nu \nabla^s \bm{u}) + (r,\nabla \cdot \bm{u}),
\end{equation}
and nonlinear advection represented by the Reynolds projection
\begin{equation}
  \mathcal{R}(\bm{w},\bm{u}) = \mathcal{B}(\bm{w},\bm{u},\bm{u}),
\end{equation}
where $\mathcal{B}$ is the tri-linear term.
\begin{equation}
  \mathcal{B}(\bm{w},\bm{u},\bm{v}) \equiv(\nabla \bm{w}, \bm{u}\bm{v}). 
\end{equation}

To take into account the multiscale representation, we write the
solution space $\mathcal{V}$ as a disjoint sum 
\[  \mathcal{V} = \overline{\mathcal{V}} \oplus
\widetilde{\mathcal{V}} \oplus \widehat{\mathcal{V}}, \] 
in which $\overline{\mathcal{V}}$ and $\widetilde{\mathcal{V}}$
comprise the large and small scales, respectively, whereas
$\widehat{\mathcal{V}}$ contains the unresolved scales that cannot be
represented by the numerical discretization. The scale partitioning is
sketched in Fig.~\ref{fig:scale-separation}.
\begin{figure}
\begin{center}
\includegraphics{fig1.eps}
\end{center}
\caption{
Schematic of the turbulent energy spectrum with scale partitioning
\label{fig:scale-separation}
}
\end{figure}
Now, by decomposing the test and trial functions in these spaces
\begin{align*}
  U &= \overline{U} + \widetilde{U} + \widehat{U}, \\
  W &= \overline{W} + \widetilde{W} + \widehat{W}, 
\end{align*}
we can develop exact variational equations governing different scales.
Furthermore, by assuming that the scale partitioning is orthogonal, we
obtain the following equations governing the large, the small, and the
unresolved scales:
\begin{subequations}
\label{eq:exact-vms-equations}
\begin{align}
\begin{split}
  \label{eq:exact-large-equations}
  &\mathcal{L}(\overline{W}, \overline{U}) -
  \mathcal{R}(\overline{\bm{w}}, \overline{\bm{u}}) 
       - (\overline{W},F) - \mathcal{R}(\overline{\bm{w}}, \widetilde{\bm{u}})
       - \mathcal{C}(\overline{\bm{w}}, \overline{\bm{u}}, \widetilde{\bm{u}})
  \\&\qquad= \mathcal{R}(\overline{\bm{w}}, \widehat{\bm{u}}) + 
                \mathcal{C}(\overline{\bm{w}}, \overline{\bm{u}},
  \widehat{\bm{u}}) 
 + \mathcal{C}(\overline{\bm{w}}, \widetilde{\bm{u}}, \widehat{\bm{u}}), 
\end{split}\\ 
\begin{split}
  \label{eq:exact-small-equations}
   &\mathcal{L}(\widetilde{W}, \widetilde{U}) -
   \mathcal{R}(\widetilde{\bm{u}}, \widetilde{\bm{u}}) 
        - (\widetilde{W},F) - \mathcal{R}(\widetilde{\bm{u}}, \overline{\bm{u}})
        - \mathcal{C}(\widetilde{\bm{u}}, \overline{\bm{u}}, \widetilde{\bm{u}})
  \\&\qquad = \mathcal{R}(\widetilde{\bm{w}}, \widehat{\bm{u}}) + 
                \mathcal{C}(\widetilde{\bm{u}}, \overline{\bm{u}},
   \widehat{\bm{u}})  
              + \mathcal{C}(\widetilde{\bm{w}}, \widetilde{\bm{u}},
   \widehat{\bm{u}}), 
\end{split}\\
\begin{split}
  \label{eq:exact-unresolved-equations}
     &\mathcal{L}(\widehat{W}, \widehat{U}) -
     \mathcal{R}(\widehat{\bm{w}}, \widehat{u}) 
          - \mathcal{C}(\widehat{\bm{w}}, \overline{\bm{u}}, \widehat{\bm{u}})
          - \mathcal{C}(\widehat{\bm{w}}, \widetilde{\bm{u}}, \widehat{\bm{u}}) 
     \\ &\qquad = \mathcal{R}(\widehat{\bm{w}}, \overline{\bm{u}}) +
     \mathcal{R}(\widehat{\bm{w}}, \widetilde{\bm{u}}) 
                + \mathcal{C}(\widehat{\bm{w}}, \overline{\bm{u}},
     \widetilde{\bm{u}}) + (\widehat{W},F), 
%     \\ = { \mathcal{R}(\widehat{w}, \overline{u}+\widetilde{u}) + (\widehat{W},F) }
\end{split}
\end{align}
\end{subequations}
where $\mathcal{C}(\bm{w},\bm{u},\bm{u}') =
\mathcal{B}(\bm{w},\bm{u},\bm{u}') +
\mathcal{B}(\bm{w},\bm{u}',\bm{u})$ is 
the cross stress term.
We have written these equations in a form such that all terms that
depend on the unresolved scales are collected in the right-hand sides.
It is thus evident that there is an effect of the unresolved scales on
the computable, resolved scales, and it goes without saying that this
effect must be modeled. In the original paper by Hughes et
al.~\cite{TJRHughes_LMazzei_KEJansen_2000a}, the modelling assumptions
were not stated, but the issue was clarified by
Collis~\cite{SSCollis_2001a}, who showed that essentially the following
assumptions result in a method that is identical to the method
proposed by Hughes (which by then had produced excellent results~%
\cite{TJRHughes_LMazzei_AAOberai_AAWray_2001a,TJRHughes_AAOberai_LMazzei_2001a})
\begin{itemize}
\item{The separation between large and unresolved scales is
      sufficiently large so that there is negligible  
      \emph{direct} dynamic influence from the unresolved scales on
      the large scales.} 
\item{The dynamic impact of the unresolved scales on the small scales
  are on average dissipative in nature.}
\end{itemize}
The simple scalar Smagorinsky-type model is in an averaged sense fully
consistent with the last assumption. In order to approximate the
temporal behaviour at the cut-off, a more refined modelling approach
would be needed. This is however outside the scope of the present
study.

With these assumptions, the LES model is only applied to the small
scale equation, adding additional dissipation where it is mostly
needed. Different implementations of this method by the Hughes group~%
\cite{TJRHughes_LMazzei_AAOberai_AAWray_2001a,TJRHughes_AAOberai_LMazzei_2001a},
by Ramakrishnan and Collis~%
\cite{SRamakrishnan_SSCollis_2002b},
and by Jeanmart and Winckelmans~%
\cite{HJeanmart_GSWinckelmans_2002a}
have produced very good results even for wall-bounded channel flows.

We remark here that both assumptions are, or at least should be, open
to scrutiny.  Firstly, although it is plausible that the unresolved
scales do not influence the large scales, it is not necessarily
obvious.  In fact, a recent analysis by Oberai \etal~%
\cite{AAOberai_VGravemeier_GCBurton_2004a} showed that the energy
transfer from the large and small scales, respectively, to the
unresolved scales depends critically on the discretization method and
the function spaces that are used to perform the scale
partitioning. Furthermore, Reynolds number effects or other aspects of
the flow physics may mandate that a more sophisticated model for the
large scales must be taken into account. Secondly, the assumption of
a one-way cascade from the small to the unresolved scales require that
flow is properly resolved, such that the cut-off is far out in the
inertial range.  This is unfortunately not always the case in LES
computations.  Such considerations are, however, outside the scope of
the present study. Our objective is to present an implementation of
the VMS LES formulation in the spectral element method.  For this
purpose the assumptions employed to date~%
\cite{TJRHughes_LMazzei_KEJansen_2000a,SSCollis_2001a} are acceptable.
At present, we merely note in passing that the VMS method presents an
excellent framework for improved modelling to address these issues.

Bearing the above in mind, we can formulate the variational modeled
equations. The effect of the unresolved scales on the large scales,
given by the right-hand side of~\eqref{eq:exact-large-equations}, is
neglected according to the first assumption, while the effect of the
unresolved scales on the small scales, given by the right-hand side
of~\eqref{eq:exact-small-equations}, is modeled by a Smagorinsky
term. The equation for the unresolved scales is naturally omitted.  
The resulting set of equations is
%\[
\begin{subequations}
\label{eq:modeled-vms-equations}
\begin{gather}
        \mathcal{L}(\overline{W}, \overline{U}) 
          - \mathcal{R}(\overline{\bm{w}}, \overline{\bm{u}})
          - \mathcal{R}(\overline{\bm{w}}, \widetilde{\bm{u}}) 
          - \mathcal{C}(\overline{\bm{w}}, \overline{\bm{u}},
        \widetilde{\bm{u}}) - (\overline{W},F) = 0, %\\ 
\\
    \begin{split}
        \mathcal{L}(\widetilde{W}, \widetilde{U}) 
            - \mathcal{R}(\widetilde{\bm{w}}, \overline{\bm{u}})
            - \mathcal{R}(\widetilde{\bm{w}}, \widetilde{\bm{u}}) 
           &- \mathcal{C}(\widetilde{\bm{w}}, \overline{\bm{u}},
        \widetilde{\bm{u}})\\ 
           &- (\widetilde{W},F) 
            = - (\nabla^s \widetilde{\bm{w}}, 2 \nu_T
                    \nabla^s \widetilde{\bm{u}}).
    \end{split}
\end{gather}
\end{subequations}

The terms that couple the different scales are evident in
\eqref{eq:modeled-vms-equations}; the small-scale equation has been
supplemented with a dissipative term that accounts for the
interactions between the small and the unresolved scales, whereas
large-scale Reynolds and cross stress projection account for the
large-scale influence on the small scales. The large-scale equation
contains a projection of the small-scale Reynolds stress onto the
large-scale to account for interaction between the small and the large
scales (i.e. back-scatter).

We are however chiefly concerned with the complete resolved solution
$\overline{\widetilde{U}}=\overline{U}+\widetilde{U}$, not with the
large and small scales per se, and adding the large- and
small-scale equations we obtain
\begin{equation}
   \label{eq:vms-combined-eqs}
  \left(\overline{\widetilde{W}},
         \mathcal{N}(\overline{\widetilde{U}})\right) +  
         (\nabla^s \widetilde{\bm{w}}, 2 \nu_T \nabla^s \widetilde{\bm{u}})  
   =  (\overline{\widetilde{W}},F).
\end{equation}
We note that in this equation, all the interaction terms between the
large and the small scales are accounted for in the advection operator
$\mathcal{R}$, which is part of the first term on the left-hand side 
in~\eqref{eq:vms-combined-eqs}. The projected cross and Reynolds
stress terms that appear in the large- and small-scale equations
\eqref{eq:modeled-vms-equations} are therefore mainly important for
analysis and turbulence modelling, but need not necessarily impact on
the implementation of the method. The variational formulation is hence
primarily an analysis tool and a vehicle for developing the VMS
methodology.  The essential feature of the method is that the
turbulence modelling should be confined to the small scales. As long as
a suitable scale partitioning can be performed on the solution space,
the methodology can in principle be applied to any discretization, as
indicated by Hughes \etal~%
\cite{TJRHughes_LMazzei_AAOberai_AAWray_2001a}.

% -----------------------------------------------------------------------------
\section{Implementation in the spectral element method}
\label{sec:sem}
In this section we describe the implementation of a VMS LES model in a
high-order spectral element method for the solution of the
incompressible Navier-Stokes equations.

We will start with a brief discussion of Legendre polynomials and the
spectral element basis functions. These concepts are important, both
for the description of the basic numerical method as well as for the
implementation of the variational multiscale framework that
follows. More details about the topics covered in
Sections~\ref{sec_Legendre} and~\ref{sec_NSsolver} can be found
in~\cite{MODeville_PFFischer_EHMund_2002a}.

\subsection{Legendre spectral elements}
\label{sec_Legendre}
The Legendre polynomials are orthogonal with respect to the 
unweighted inner product in the function
space $L^2(-1,1)$. 
The Legendre polynomials are given by the
recurrence relation  
\begin{equation}
\begin{aligned}
  L_0(x) &= 1, \\ 
  L_1(x) &= x, \\ 
  L_{k+1}(x) &= \frac{2k+1}{k+1} x L_k(x) - \frac{k}{k+1} L_{k-1}(x),
  \quad k \ge 1, 
\end{aligned}
\label{eq:legendre-polys}
\end{equation}
where $L_N(x)$ is the Legendre polynomial of degree $N$. 

The Gauss-Lobatto-Legendre (GLL) points
$\left\{ \xi_j \right\}_{j=0}^{N}$ on $\Lambda = [-1,1]$ are defined
as the extrema of the $N$th order Legendre polynomial $L_N(x)$, in
addition to the endpoints of $\Lambda$: 
\begin{equation} \label{GLLpoints}
  \xi_0 = -1, \, \xi_N = 1, \quad L_N'(\xi_j) = 0, \,
  j=1,\ldots,N-1, \, \xi_0 < \xi_1 < \ldots < \xi_N.
\end{equation} 
Furthermore, the Gauss-Legendre (GL) points 
$\left\{ \eta_j \right\}_{j=1}^{N-1}$ on $\Lambda$,
that are used to represent the pressure in the spectral element method,
are defined implicitly by  $L_{N-1}(\eta_j)=0$, i.e. as the 
zeros of the Legendre polynomials of order $(N-1)$%
~\cite{canuto-etal:spectral-98}. 
Note that the GL points do not include the
endpoints of $\Lambda$. 

The spectral element nodal Gauss-Lobatto-Legendre basis is
defined by choosing trial and test functions to be the corresponding
Lagrangian interpolants at the Gauss-Lobatto-Legendre (GLL) grid
points, constructed as $N$th order polynomials such that each
function has the property
\begin{equation} \label{Lagrangedelta}
  h_j(\xi_i) = \delta_{ij}, \quad i = 0,\ldots,N.
\end{equation}
A function $w(x)$ defined on $\Lambda$ can then be represented 
by the interpolating polynomial:
\begin{equation} \label{elementbasis}
  w_h(x) = \sum_{i=0}^{N} w_i h_i(x), \qquad x \in \Lambda,
\end{equation}
where $w_i = w(\xi_{i})$ are the function values at the GLL
points. Higher-dimen\-sional trial and test functions are constructed as
tensor products of these one-dimen\-sional functions. Each velocity
component is represented this way on each element, and the global
representation is the sum of the representations on all elements.

A Gauss-Legendre nodal basis for the pressure is constructed in an
analogous way, only taking into account that we use lower-order
polynomials in the basis for the pressure to avoid spurious pressure
modes in the solution~\cite{YMaday_ATPatera_1989a}.

\subsection{Spectral element Navier-Stokes solver}
\label{sec_NSsolver}
To solve the Navier-Stokes equations~\eqref{eq:navier-stokes} we
employ an implicit-explicit time splitting in which we integrate the
advective term explicitly, while we treat the diffusive term, the
pressure term, and the divergence equation implicitly. After
discretization in time we can write~\eqref{eq:navier-stokes} in
the form
\begin{subequations}
\label{eq:semi-discrete-ns}
\begin{align}
(\alpha I - \nu\nabla^2) \bm{u}^{n+1} &= \nabla p + \bm{g}(\bm{f},
  \bm{u}^{n}, \bm{u}^{n-1}, \dots),\\ 
\nabla\cdot \bm{u}^{n+1} &= 0,
\end{align}
\end{subequations}
in which the explicit treatment of the advection term is included in
the source term $\bm{g}$.  In the actual implementation we use the
BDF2 formula for the transient term,
\[
  \frac{\partial \bm{u}}{\partial t} = 
  \frac{3\bm{u}^{n+1} -4\bm{u}^{n}+\bm{u}^{n-1}}{2\Delta t} + 
  O(\Delta t^2),
\]
which gives $\alpha=3/2\Delta t$ in~\eqref{eq:semi-discrete-ns},
while we compute the advective contributions according to the 
operator-integration-factor (OIF) method~%
\cite{YMaday_APatera_EMRonquist_1990a}. 

The spatial discretization is based on a spectral element method~%
\cite{YMaday_ATPatera_1989a,ATPatera_1984a}; the computational domain
is sub-divided into non-overlapping hexahedral cells or elements.
Within each element, a weak representation
of~\eqref{eq:semi-discrete-ns} is discretized by a Galerkin method in
which we choose the test and trial functions from bases of polynomial
spaces
\begin{subequations}
\label{eq:polspaces}
\begin{align}
u^{h}_{i} &\in P_{N}(x) \otimes P_{N}(y) \otimes P_{N}(z), \\
p^{h} &\in P_{N-2}(x) \otimes P_{N-2}(y) \otimes P_{N-2}(z).
\end{align}
\end{subequations}
The velocity component variables are defined in the
Gauss-Legendre-Lobatto basis described above, and they are
$C^{0}$-continuous across element boundaries.  The pressure variable
is represented in the Gauss-Legendre basis, and is discontinuous
across element boundaries.  As we noted above, the unknowns, or
dependent variables, in the discrete formulation are the function
values of the velocities in the GLL points, and of the pressure in the
GL points.

The GLL grid corresponding to the Legendre polynomial of degree $N$
has ($N+1$) points. Gauss-Lobatto-Legendre quadrature at the ($N+1$)
GLL points is exact for polynomials of degree up to ($2N-1$). Hence,
the computation of the inner products corresponding to the diffusive
terms in~\eqref{eq:navier-stokes} are calculated exactly, whereas the
evaluation of the non-linear advective terms incurs quadrature
(aliasing) errors. These errors can be detrimental to the stability of
the method and must be controlled. The most fundamental approach to
de-aliasing is to perform over-integration~%
\cite{PFFischer_2002a,RMKirby_GEKarniadakis_2003a} -- that is, to
over-sample by a factor 3/2 and calculate the quadrature at this
refined grid for the inner products containing non-linear terms. 
The overhead involved depends on the amount of the total computational
time that is originally spent on the advection part, but for the
channel flow calculations presented here, over-integration typically
leads to an increase of around 20\%\ of computational time.
 
An alternative, and computationally more efficient approach, is to use
polynomial filtering of the solutions as proposed by Fischer and Mullen~%
\cite{PFFischer_JSMullen_2001a}, where a simple filter operator with
negligible computational cost is applied to the solution at every
time-step. The effect in the spectral space on each element is to
transfer a certain fraction (the filter strength) of the
energy on the highest order basis polynomial in each element over to the
third-highest order polynomial~\cite{RPasquetti_CJXu_2002}. By this
operation, the pile-up of energy on the highest order polynomial is
reduced, while the values at the element boundaries are
unchanged. Filter strengths as small as 1--5\%\ can have positive
effects on the solution.

For the solution of the discrete system of equations we now introduce the
discrete Helmholtz operator,
\[H = \frac{3}{2\Delta t} B + \nu A, \] 
where $A$ and $B$ are the global three-dimensional stiffness- and mass
matrices; the discrete divergence operator, $D$; and the discrete
gradient operator, $G$. Appropriate boundary conditions should be
included in these discrete operators. This gives the discrete
equations
\begin{subequations}
\label{eq:fully-discrete-ns}
\begin{align} 
  H\bm{u}^{n+1} - Gp^{n+1}  &= B\bm{f}, \label{eq:fully-discrete-momentum}\\ 
  -D\bm{u}^{n+1}            &= 0, \label{eq:fully-discrete-continuity}
\end{align}
\end{subequations}
where the change of sign in the pressure gradient term is caused by an
integration by parts in the construction of the weak form of the
problem.  This discrete system is solved efficiently by a second order
accurate pressure correction method.  If we let $Q$ denote an
approximate inverse to the Helmholtz operator, given by a scaled
inverse of the diagonal mass matrix, the pressure correction method
can be written~%
\begin{subequations}
\label{eq:pressure-correction-method}
\begin{align} 
    H\bm{u}^* & =  B\bm{f} + Gp^n, \\
    DQG(p^{n+1}-p^n) & = - D\bm{u}^*  \\  
    \bm{u}^{n+1} & =  \bm{u}^* + QG(p^{n+1} - p^n),
\end{align}
\end{subequations}
where $u^*$ is an auxiliary velocity field that does not satisfy the 
continuity equation, i.e.~$D\bm{u}^* \neq 0$.

The discrete Helmholtz operator is symmetric and diagonally dominant, 
since the mass matrix of the Legendre discretization is diagonal, 
and can be efficiently solved by the conjugate gradient method with a
diagonal (Jacobi) preconditioner.
Whereas the pressure operator $DQG$ is easily computed; it is 
ill-conditioned. The pressure system is solved by the
preconditioned conjugate gradient method, with a multilevel overlapping Schwarz
preconditioner based on linear finite elements%
~\cite{PFFischer_NIMiller_HMTufo_2000a}.

\subsection{Incorporation of VMS LES in the SEM}
The implementations of the variational multiscale LES method reported
in~% 
\cite{TJRHughes_LMazzei_AAOberai_AAWray_2001a,TJRHughes_AAOberai_LMazzei_2001a,SRamakrishnan_SSCollis_2002b}
used global spectral methods. These methods naturally employ an
orthogonal modal basis, such that the scale partitioning becomes
straightforward. Our spectral element code uses on an element-wise
discretization based on the Legendre polynomials.  The Legendre
polynomials offer an orthogonal hierarchical basis. Like the majority
of the spectral element community, we do however use a \emph{nodal}
basis constructed from the Lagrangian interpolant functions. In this
case all the basis functions contains information on all the scales
and the scale partitioning is no longer straightforward.

\subsubsection{Nodal and modal bases}
\label{sec:nodal-modal}
We have demonstrated above that in the \emph{nodal}
Gauss-Lobatto-Legendre basis a function $w(x)$ defined on $-1\leq x
\leq 1$ can be represented by a combination of the interpolating
polynomials, as given by~\eqref{elementbasis}. The coefficients in the
sum are the function values at the grid points.

An alternative, \emph{modal}, representation is to use an expansion
directly in the Legendre polynomials
\begin{equation}
w(\xi) = \sum_{j=0}^{N} \, c_j \, \sqrt{\frac{2j+1}{2}}\, L_j(\xi),
\end{equation}
where the unknowns now are the spectral coefficients $c_j$. The factor
$\sqrt{\frac{2j+1}{2}}$ 
is used to normalize the basis. The scaled Legendre
polynomials represents a natural orthonormal basis, in which it is
straightforward to perform the scale partitioning.  In this setting,
it is natural to associate the low order polynomials with the large
scales and the higher order polynomials with the smaller scales.

The nodal and modal bases are related through the linear transformation
\begin{equation}
{K}\bm{c} = \bm{w},
\end{equation}
where the entries of the matrix ${K}$ are given by
\[
({K})_{ij} = \sqrt{\frac{2j+1}{2}} L_j(\xi_i),
\]
and $\bm{c}$ and $\bm{w}$ are the vectors of spectral coefficients
and GLL nodal function values, respectively.

Let $N=\overline{N} + \widetilde{N}$, such that $\overline{N}$ is the
dimension of the 
polynomial basis for the large scales and $\widetilde{N}$ is the dimension of
the small-scale space. The large-scale part of a nodal function $\bm{w}$ 
can then be written as
\begin{equation}
\widetilde{\bm{w}} = {KTK}^{-1}\bm{w},
\label{eq:large-scale-projection}
\end{equation}
where ${T}=\mathrm{diag}({I}_{\widetilde{N}},{0}_{\overline{N}})$ is the
operator that annihilates the small-scale components in the modal 
basis. For notational convenience, we define the large-scale extraction
operator
\[ L = KTK^{-1}, \]
while the corresponding small-scale extraction operator is
\[ S = I - L. \]

\begin{figure}
\begin{center}
\includegraphics[scale=0.45,angle=0]{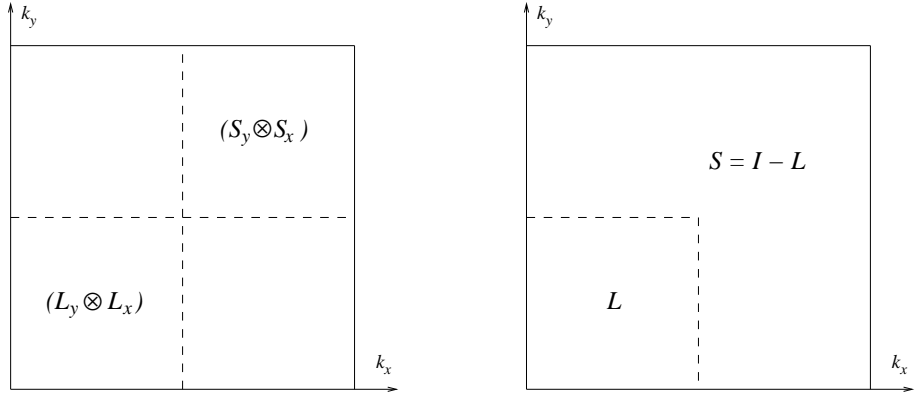}
\end{center}
\caption{Large- and small-scale partitions in the 2-dimensional
  polynomial wavenumber space. The chosen partition operators are
  shown to the right.}
\label{fig:filter2d}
\end{figure}

When tensor products of these operators are formed in higher
dimensions, the resulting operators extract the components with large-scale,
or small-scale, respectively, components in {\em all}
dimensions. The sum of these two operators does not add up to the
identity, so we choose to define the three-dimensional small-scale
extraction operator to be
\begin{equation}
  S = I - (L_z \otimes L_y \otimes L_x).
\end{equation}
This is illustrated in two dimensions in
figure~\ref{fig:filter2d}. The resulting small-scale extraction
operator returns functions with small-scale structure in {\em at least
  one} dimension.

\subsubsection{Properties of the large-small partition}

\begin{figure}
\begin{center}
\includegraphics[scale=1,angle=0]{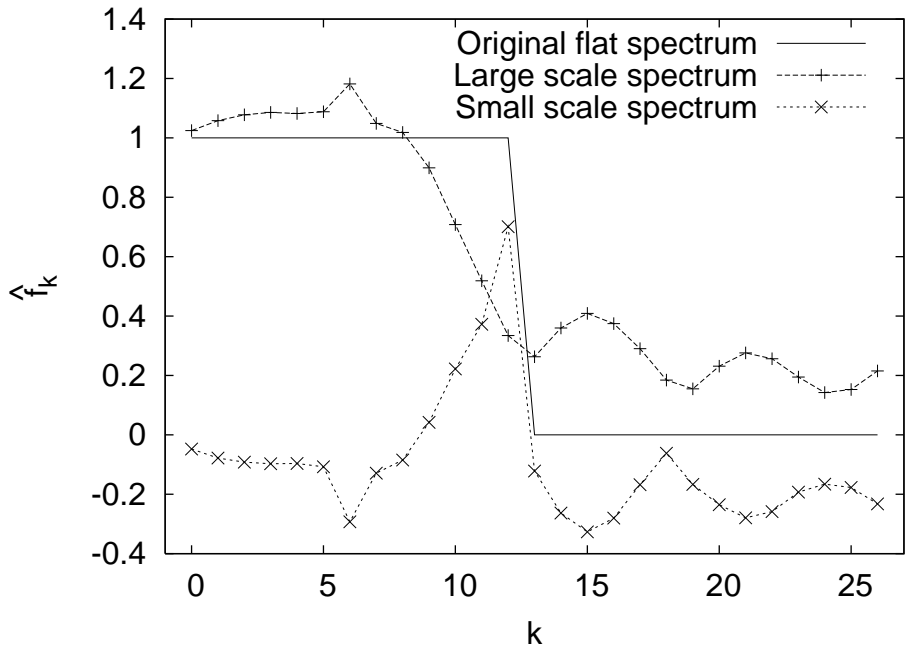}
\end{center}
\caption{Fourier representation of a sharp cut-off in Legendre
  modal space.}
\label{fig:scalepart}
\end{figure}

The large-scale extraction operator corresponds to a sharp cut-off in
the Legendre modal space. To illustrate the effect in Fourier space of
this large-small partitioning, we represent the function
\begin{equation}
  f(x) = \sum_{k=0}^{12}\, \cos(kx)
\end{equation}
on a spectral element grid on $[0,2\pi]$ with 6 elements and 7 grid
points in each element. Higher wave-numbers can not be represented
accurately on this particular grid. We extract the large- and
small-scale partitions using $\overline{N}=4$ of the 7 modes (57\%) on
each element as the large-scale space. The two resulting functions are
sampled on a 54-point regular grid, and their Fourier spectra
are plotted in figure~\ref{fig:scalepart}.

The main point illustrated by figure~\ref{fig:scalepart} is that
although the scale partitioning in the Legendre space is done as a
sharp cut-off, the Fourier spectra of the two partitions are much
smoother. The reason for this is that each the original cosine terms
is represented by a combination of local Legendre modes on each
element. We also note that the gradual growth in the small-scale
spectrum starts around the cut-off percentage, so the impact of the
small-scale extraction is weaker than for a straightforward Fourier
representation.

In a more general case with variable element size and/or polynomial
order, it may be possible to vary the cut-off point in the local
Legendre space to keep the corresponding ``average wavelength''
approximately constant throughout the whole domain.

\subsubsection{Implementation of the model term}
We now turn our attention to the implementation of the variational
multiscale model term 
$(\nabla^s\widetilde{\bm{w}}, 2\nu_{T}\nabla^s\widetilde{\bm{u}})$ 
from \eqref{eq:vms-combined-eqs}.
Note that the turbulent eddy viscosity $\nu_T$ is not a material
property of the fluid, but a property of the flow field and as such
varies through the flow domain.

It is instructive to first consider the one-dimensional case with a
constant eddy viscosity.
Furthermore, for ease of exposition, we only consider a single element.
In this case, the weak form of the model term above is
\begin{equation}
  \nu_T \int \frac{\partial\widetilde{w}}{\partial x}
  \frac{\partial\widetilde{u}}{\partial x}.  
\end{equation}
Using the small-scale extraction operator defined above, we have
\begin{equation}
  \widetilde{u}(x) = \sum_{m=0}^{N} \sum_{q=0}^{N} S_{mq} u^q h_m(x),
\end{equation}
and for a given test function on Lagrange form ($w^i(x) = h_i(x)$)
\begin{equation}
  \widetilde{w}^i(x) = \sum_{p=0}^{N} S_{pi} h_p(x). 
\end{equation}
Inserting these representations and using Gauss-Lobatto quadrature, we obtain 
\begin{equation} \label{eq:1D-small-scale-stiffness}
\begin{split} 
  (\nabla \widetilde{w}, \nabla \widetilde{u}) & =\sum_{r=0}^{N}
  \sum_{p=0}^{N} \sum_{m=0}^{N} \sum_{q=0}^{N} 
  S_{pi} h'_p(\xi_r) S_{mq} u^q h'_m(\xi_r) \rho_r \\
  & = \sum_{p=0}^{N} S_{pi} \sum_{m=0}^{N}
  \sum_{q=0}^{N} S_{mq} u^q \sum_{r=0}^{N} h'_p(\xi_r) 
  h'_m(\xi_r) \rho_r \\
  & = \sum_{p=0}^{N} S_{pi} \sum_{m=0}^{N}
  \sum_{q=0}^{N} S_{mq} u^q A_{pm} \\
  & = \sum_{q=0}^{N} (S^T A S)_{iq} u^q = S^T A S u \\
  & = (I-L)^T A \widetilde{u},
\end{split}
\end{equation}
where the final line is in the form we generalize to higher dimensions. 
It is easily seen from the second-to-last line
that~\eqref{eq:1D-small-scale-stiffness} represents a symmetric
operator acting on $u$.

The corresponding term in three dimensions is
\begin{equation} \label{eq:3D-const-small-scale-stiffness}
\begin{split} 
  \left( \nabla \widetilde{w}, \nabla \widetilde{u}_i \right)
  & = \left( (B^z \otimes B^y \otimes A^x) - (L^{zT} \otimes L^{yT}
  \otimes L^{xT})(B^z \otimes B^y \otimes A^x) \right) \widetilde{u}_i \\ 
  & + \left( (B^z \otimes A^y \otimes B^x) - (L^{zT} \otimes L^{yT}
  \otimes L^{xT})(B^z \otimes A^y \otimes B^x) \right) \widetilde{u}_i \\ 
  & + \left( (A^z \otimes B^y \otimes B^x) - (L^{zT} \otimes L^{yT}
  \otimes L^{xT})(A^z \otimes B^y \otimes B^x) \right) \widetilde{u}_i
\end{split}
\end{equation}
for each component $u_i$.

Taking into account that the eddy viscosity, $\nu_T(x,y)$, is not
constant but rather a function that varies in space, will distort the
tensor product structure of 
\eqref{eq:3D-const-small-scale-stiffness}.  
Following the procedure for discretization of terms with variable coefficients
described in~%
\cite{MODeville_PFFischer_EHMund_2002a}, 
we can write 
\begin{equation}
  \label{eq:3d-var-small-scale-stiffness}
\begin{split} 
  \left( \nabla \widetilde{w},  2\nu_T(x,y,z)  \nabla \widetilde{u}_i \right) 
  & = 2 \, \left( I^{z} \otimes I^{y} \otimes D^{xT} \right) {V} \left( I^z
  \otimes I^y \otimes D^x \right) \widetilde{u}_i \\
  &\quad - 2 \, \left( L^{zT} \otimes L^{yT} \otimes (D^{x} L^{x})^T
  \right) {V} \left( I^z \otimes I^y \otimes D^x \right)
  \widetilde{u}_i \\ 
  & + 2 \, \left( I^{z} \otimes D^{yT} \otimes I^x \right) {V} \left( I^z
  \otimes D^y \otimes I^x \right) \widetilde{u}_i \\
  & \quad - 2 \, \left( L^{zT} \otimes (D^{y} L^{y})^T \otimes L^{xT}
  \right) {V} \left( I^z \otimes D^y \otimes I^x \right)
  \widetilde{u}_i \\
  & + 2 \, \left( D^{zT} \otimes I^{y} \otimes I^x \right) {V} \left( D^z
  \otimes I^y \otimes I^x \right) \widetilde{u}_i \\
  & \quad - 2 \, \left( (D^{z} L^{z})^T \otimes L^{yT} \otimes L^{xT}
  \right) {V} \left( D^z \otimes I^y \otimes I^x \right)
  \widetilde{u}_i.  
\end{split} 
\end{equation}
In this equation, $D$ denotes the GLL derivation matrix in each
direction.  Furthermore, the values of the eddy viscosity are lumped
with the GLL integration weights in the diagonal matrix $V$ with the
entries $\nu_T^{rst} \rho^x_r \rho^y_s \rho^z_t$, in which $rst$ are
the grid point indices and $V$ is ordered to be consistent with the
ordering of the element grid points.

We are now finally ready to consider the model term in the form given
in~\eqref{eq:vms-combined-eqs}. Since the product of a symmetric and an 
anti-symmetric tensor is zero, we find that we only 
need to to compute the inner product
\begin{equation} \label{eq:3d-vms-dissipation-term}
  \left( \nabla \widetilde{w}, 2  \nu_T  \nabla^s \widetilde{\bm{u}} \right)
  = \left( \frac{\partial \widetilde{w}}{\partial x_j}, 
           \nu_T  \frac{\partial \widetilde{{u}}_i}{\partial x_j} \right) +
    \left( \frac{\partial \widetilde{w}}{\partial x_j}, 
           \nu_T  \frac{\partial \widetilde{{u}}_j}{\partial x_i} \right).
\end{equation}
In tensor product form, the VMS small-scale dissipation term for the
component of the momentum equation becomes
\begin{equation} \label{eq:full-3d-vms-dissipation-term}
\begin{array}{l}
  \left( \nabla \widetilde{w},  2\nu_T(x,y,z)  \nabla \widetilde{u}_1
  \right) \\
  \quad  = 2 \, \left\{ \left( I^{z} \otimes I^{y} \otimes D^{xT} 
  \right) -  \left( L^{zT} \otimes L^{yT} \otimes (D^{x} L^{x})^T
  \right) \right\} {V} \left( I^z \otimes I^y \otimes D^x \right)
  \widetilde{u}_1 \\ 

  \qquad  + \, \left\{ \left( I^{z} \otimes D^{yT} \otimes I^{x}
  \right) -  \left( L^{zT} \otimes (D^{y} L^{y})^T \otimes L^{xT}
  \right) \right\} {V} \left( I^z \otimes D^y \otimes I^x \right)
  \widetilde{u}_1 \\ 

  \qquad  + \, \left\{ \left( D^{zT} \otimes I^{y} \otimes I^{x}
  \right) -  \left( (D^{z} L^{z})^T  \otimes L^{yT} \otimes L^{xT}
  \right) \right\} {V} \left( D^z \otimes I^y \otimes I^x \right)
  \widetilde{u}_1 \\ 

  \qquad  + \, \left\{ \left( I^{z} \otimes D^{yT} \otimes I^{x}
  \right) -  \left( L^{zT} \otimes (D^{y} L^{y})^T \otimes L^{xT}
  \right) \right\} {V} \left( I^z \otimes I^y \otimes D^x \right)
  \widetilde{u}_2 \\ 

  \qquad  + \, \left\{ \left(  D^{zT}  \otimes I^{y} \otimes I^{x}
  \right) -  \left( (D^{z} L^{z})^T  \otimes L^{yT} \otimes L^{xT}
  \right) \right\} {V} \left( I^z \otimes I^y \otimes D^x \right)
  \widetilde{u}_3,   
\end{array} 
\end{equation}
and we obtain similar expressions for the other two components.  
The couplings between the velocity components, introduced by the
second term of~\eqref{eq:3d-vms-dissipation-term}, are handled by
including the cross terms in the explicit part of the time splitting,
leaving the Helmholtz problem for the velocity components uncoupled.  

As seen from~\eqref{eq:full-3d-vms-dissipation-term}, the calculation of
the VMS LES model terms requires several additional operations. The
increase in total computational work will vary with the size and
complexity of the simulation, but for the cases considered in this
paper the increase is in the range 20--40\%, with the smallest
relative increase for the largest simulations. To put these numbers
into perspective, we note that the total computational complexity of
the spectral element method is $O(K^3N^4)$, so increasing the
polynomial order ($N-1$) from 6 to 7 gives a 70\%\ increase in
computational time, about the same as increasing the number of
elements in each dimension ($K$) from 5 to 6 would give.

\subsubsection{Smagorinsky model}

The eddy viscosity $\nu_T({\bm x},t)$ is chosen
in~\cite{TJRHughes_LMazzei_KEJansen_2000a} as a Smagorinsky-type function:
\begin{equation} \label{eq:small-small-smagorinsky}
  \nu_T = (C'_S\Delta')^2 |\nabla^s \widetilde{\bm u}|,
\end{equation}
or alternatively
\begin{equation} \label{eq:large-small-smagorinsky}
  \nu_T = (C'_S\Delta')^2 |\nabla^s \bar{\bm u}|.
\end{equation}
The former was labeled ``small-small'' in~%
\cite{TJRHughes_LMazzei_AAOberai_AAWray_2001a}, while
the latter was labeled ``large-small''.

As the purpose of the model term is to approximate the effect of the
unresolved scales on the small scales, it is argued
in~\cite{TJRHughes_LMazzei_KEJansen_2000a}
that~\eqref{eq:small-small-smagorinsky} is more consistent with the
physical basis of the method, whereas~\eqref{eq:large-small-smagorinsky}
appears to be a computationally attractive alternative. The results
in~\cite{TJRHughes_LMazzei_AAOberai_AAWray_2001a,TJRHughes_AAOberai_LMazzei_2001a}
show that good results are obtained with both methods. However, in
terms of the spectral element implementation, the ``large-small'' form
is not a computational simplification. A more attractive form is
instead the ``full-small'' term
\begin{equation} \label{eq:full-small-smagorinsky}
  \nu_T = (C'_S\Delta')^2 |\nabla^s {\bm u}|,
\end{equation}
in which the scale extraction operators are avoided completely.

The sum $|\nabla^s {\bm u}|$ can be written out as
\begin{equation}
  |\nabla^s {\bm u}| = \sqrt{ \frac{1}{2} \,  \sum_{i=1}^{3}
  \sum_{j=1}^{3} \left( \frac{\partial u_i}{\partial x_j} +
  \frac{\partial u_j}{\partial x_i} \right)^2 }.
\end{equation} 
The constant $C'_S$ is set to $0.1$,
following~%
\cite{TJRHughes_LMazzei_AAOberai_AAWray_2001a,TJRHughes_AAOberai_LMazzei_2001a},
while $\Delta'$ is
calculated for each element as the geometric average of the mean grid
spacing in each direction.

% -----------------------------------------------------------------------------
\section{Computational results}
\label{sec:results}
\subsection{Channel flow}
The plane turbulent channel flow is one of the simplest cases of
an inhomogeneous turbulence field, and this configuration has therefore
been extensively used to assess the performance of turbulence models.
The fully developed, statistically stationary, plane channel flow
is an equilibrium flow, because there is a global balance between 
the rate of production of turbulent kinetic energy and the rate of 
viscous dissipation. 

The fluid domain is bounded by two infinitely large parallel solid walls,
and the flow is driven by a constant mean pressure gradient in the stream-wise
direction along the walls.
The boundary conditions are no-slip at the solid walls, and 
periodicity is imposed in the streamwise ($x$) and spanwise ($z$)
directions, respectively. The wall-normal direction is thus $y$, and
the channel \emph{half-height} is denoted $h$.

The instantaneous flow field is three-dimensional and
time dependent, the ensemble (or time) averaged flow field is however
unidirectional. If we let $\langle\cdot\rangle$ denote the ensemble
average, we therefore have $\mathbf{U} = \langle\bm{u}\rangle  =
[U(y), 0, 0]$. 

The \emph{friction velocity}, $u_\tau$, is defined by
\begin{equation}
  u_\tau^2 \equiv \nu \cdot \left. \frac{dU}{dy} \right|_{\mathrm{wall}},
\end{equation}
and this is used in the definition of the relevant Reynolds number for
plane channel flow: $\mathrm{Re}_\tau \equiv u_\tau h / \nu$.

Integrating the ensemble averaged Navier-Stokes equations in the
wall-normal direction yields 
\begin{equation}
  0 = -\left(\frac{dP}{dx}\right) y + \mu \frac{dU}{dy} - \rho
  \langle u'v'\rangle, 
\end{equation}
where the pressure gradient is a constant, related to the Reynolds
number by
\begin{equation}
  -\left(\frac{dP}{dx}\right) = \frac{\mu^2}{h^3}\,\mathrm{Re}^2_\tau.
\end{equation}
Hence, the sum of the viscous ($\mu dU/dy$) and turbulent
($-\rho \langle u'v'\rangle$) stresses must vary linearly across the channel. 
The turbulent stress contribution dominates across the channel except
very close to the wall where viscous stress dominates. This region is
usually referred to as the viscous sub-layer and its thickness
decreases with increasing Reynolds numbers.

We consider three different Reynolds numbers: $\mathrm{Re}_\tau = 180,
\ 550, \ 950$, and the VMS LES computations are compared with
reference solutions obtained from direct numerical simulations.

\renewcommand{\arraystretch}{1.0}
\begin{table}
\begin{center}
\begin{tabular}{l|cccc}
  & Present & Moser & del \'{A}lamo & del \'{A}lamo \\  
  & DNS & \etal & \&\ Jim\'{e}nez & \etal \\ \hline 
  $\mathrm{Re}_\tau$ nom. &  180 & 180 & 550 & 950 \\
  $\mathrm{Re}_\tau$ act. & 178.83 & 178.13 & 546.75 & 934 \\
  $L_x$ & 8 & $4\pi$ & $8\pi$ & $8\pi$\\
  $L_y$ & 2 & 2 & 2 & 2 \\
  $L_z$ & 4 & $\frac{4}{3}\pi$ & $4\pi$ & $3\pi$ \\
  $N_x$ & 112 & 128 & 1536 & 3072 \\
  $N_y$ & 113 & 129 & 257 & 385 \\
  $N_z$ & 112 & 128 & 1536 & 2304 \\
  $\Delta x^+$ mean & 12.9 & 17.7 & 9.0 & 8.9 \\
  $\Delta y^+$ min & 0.10 & 0.054 & 0.041 & 0.032 \\
  $\Delta y^+$ max & 8.6 & 4.4 & 6.7 & 7.8 \\
  $\Delta z^+$ mean & 6.4 & 5.9 & 4.5 & 4.5 \\
  Elements & $16^3$ & - & - & - \\
  Pol.\ order & 7 & - & - & - \\
\end{tabular}
\caption{Grid parameters for the present DNS and the reference
  simulations by Moser \etal~\cite{RDMoser_JKim_NNMansour_1999a} and
  by del \'{A}lamo et
  al.~\cite{JCdelAlamo_JJimenez_2003,JCdelAlamo_etal_2004a}.   
  Grid spacing in wall units are calculated from the nominal
  $\mathrm{Re}_\tau$.}
\label{table:refgrids}
\end{center}
\end{table}

\renewcommand{\arraystretch}{1.0}
\begin{table}
\begin{center}
\begin{tabular}{l|cccc}
  & Coarse-24 & Coarse-36 & Coarse-42 & Coarse-60 \\ \hline 
  $\mathrm{Re}_\tau$ nom. & 180 & 180 & 550 & 950 \\
  $L_x$ & 8 & 8 & 8 & 8 \\
  $L_y$ & 2 & 2 & 2 & 2 \\
  $L_z$ & 4 & 4 & 4 & 4 \\
  $N_x$ & 24 & 36 & 42 & 60 \\
  $N_y$ & 25 & 37 & 43 & 61 \\
  $N_z$ & 24 & 36 & 42 & 60 \\
  $\Delta x^+$ mean & 40.0 & 60.0 & 104.8 & 126.7 \\
  $\Delta y^+$ min & 2.0 & 4.5 & 4.6 & 3.9 \\
  $\Delta y^+$ max & 21.1 & 29.8 & 57.4 & 68.8 \\
  $\Delta z^+$ mean & 20.0 & 30.0 & 52.4 & 63.3 \\
  Elements & $4^3$ & $6^3$ & $7^3$ & $10^3$ \\
  Pol.\ order & 6 & 6 & 6 & 6 \\
\end{tabular}
\caption{Grid parameters for the VMS LES runs. Grid spacing in wall
  units are calculated from the nominal $\mathrm{Re}_\tau$.}
\label{table:vmsgrids}
\end{center}
\end{table}

\begin{figure}
    \includegraphics[scale=1.05,angle=0]{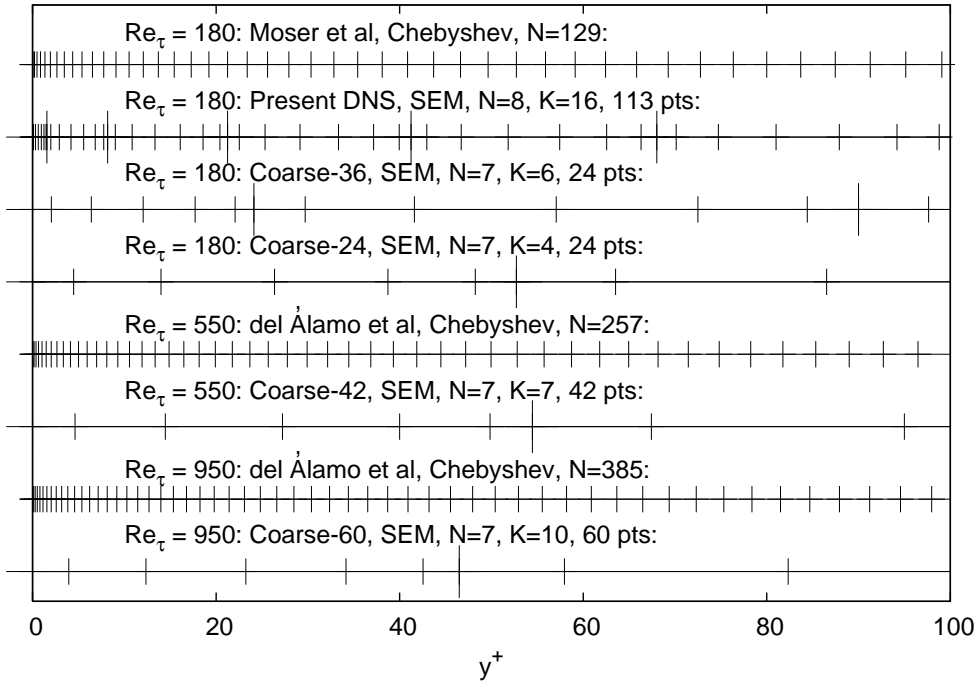} 
    \caption{Details of the point and element distribution in the
            wall-normal direction for the grids listed in
            Tables~\protect\ref{table:refgrids}
            and~\protect\ref{table:vmsgrids}. The longer bars show
            element boundaries for the spectral element grids.}  
    \label{fig:ygrids}
\end{figure}

\subsection{Direct numerical simulations at $\mathrm{Re}_\tau=180$}

As a first step towards our ultimate goal, to implement and evaluate
the variational multiscale LES method in a high order spectral element
flow solver, we performed a Direct Numerical Simulation to verify the
code.  To this end, we considered fully developed channel flow at
$\mathrm{Re}_\tau = 180$, which corresponds to the well-known
benchmark simulations reported by Kim \etal~%
\cite{kim_moin_moser_1987}. We performed the actual comparison of the
results with the updated data set reported by Moser \etal~%
\cite{RDMoser_JKim_NNMansour_1999a} who used a fully spectral
Fourier/Chebyshev method with $128\times 129\times 128$ grid points.

The simulation was carried out on a computational domain that
approximately corresponds to the one used by the reference solutions~%
\cite{RDMoser_JKim_NNMansour_1999a,kim_moin_moser_1987}, see
Table~\ref{table:refgrids} for details.  Across the channel we used 16
non-uniformly distributed elements with 8 nodal points in each
element. In the streamwise and spanwise directions we used $16\times
16$ uniformly distributed elements with $8\times 8$ nodal points per
element.  Thus, the total number of nodal points amounts to $112\times
113\times 112$ in the streamwise, wall-normal, and spanwise
directions, respectively.  The solution was advanced in time with a
time-step corresponding to 0.18 viscous time-units ($\nu/u_\tau^2$),
and with 50\%\ polynomial filtering~\cite{PFFischer_JSMullen_2001a} on
each time-step.  The simulation was initiated by a flow field
obtained from an existing plane channel flow solution obtained by a
finite-volume code.  The flow then evolved approximately 54
flow-through times before a fully developed state was achieved. The
results presented here was obtained by collecting statistics over
approx.\ 20 flow-through times.  The flow statistics are averaged over
the homogeneous -- streamwise and spanwise -- directions.  Homogeneity
in a specific direction implies that any correlation of a fluctuating
quantities remains invariant under translation in that direction.

\subsubsection{Results}
\label{dns-results}

\begin{figure}
     \includegraphics[scale=1,angle=0]{fig5.eps}
     \caption{$\mathrm{Re}_\tau = 180$: Variation of the mean velocity
    across half the channel in viscous units, compared with the 
    reference solution of Moser 
    \etal~\protect{\cite{RDMoser_JKim_NNMansour_1999a}}.} 
   \label{fig:U_log}
\end{figure}

\begin{figure}
     \includegraphics[scale=1,angle=0]{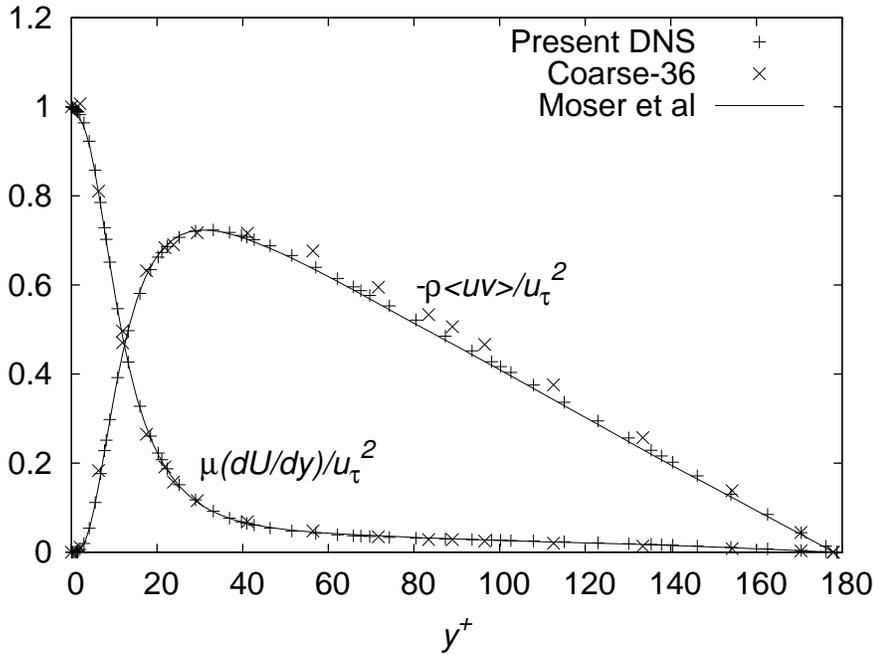}
     \caption{$\mathrm{Re}_\tau = 180$: Variation of mean viscous
            shear and the turbulent shear stress  
            across half the channel, compared with the reference solution of 
            Moser \etal~\protect{\cite{RDMoser_JKim_NNMansour_1999a}}.}
    \label{fig:shear_stress} 
\end{figure}

\begin{figure}
    \includegraphics[scale=1,angle=0]{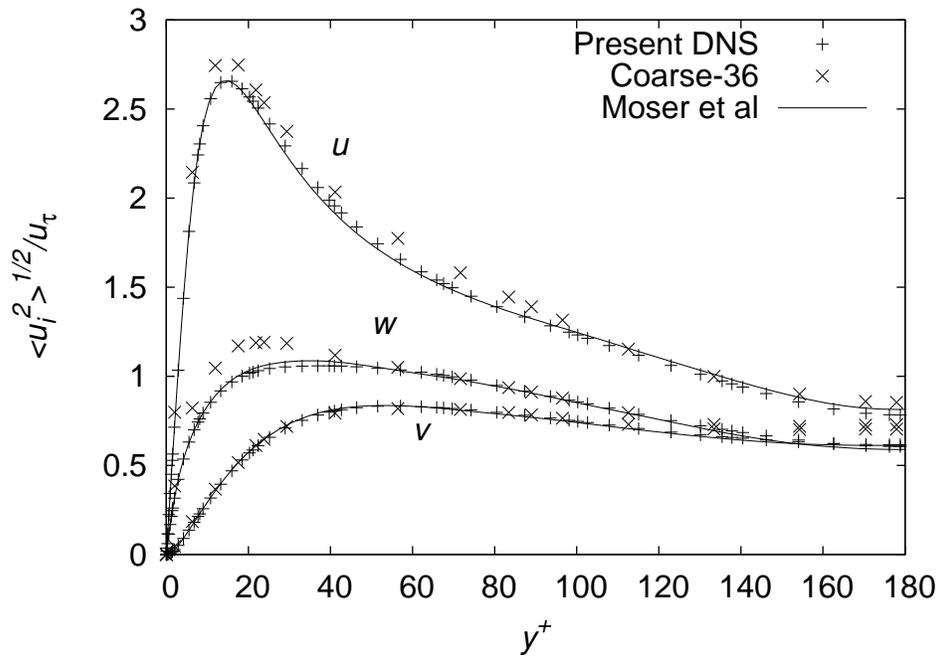} 
    \caption{$\mathrm{Re}_\tau = 180$: Variation of streamwise ($u'$),
            spanwise ($v'$), and wall-normal ($w'$)  
            root-mean-square velocity fluctuations 
            across half the channel, compared with the reference solution of 
            Moser \etal~\protect{\cite{RDMoser_JKim_NNMansour_1999a}}.}
    \label{fig:urms}
\end{figure}

\begin{figure}
     \includegraphics[scale=1,angle=0]{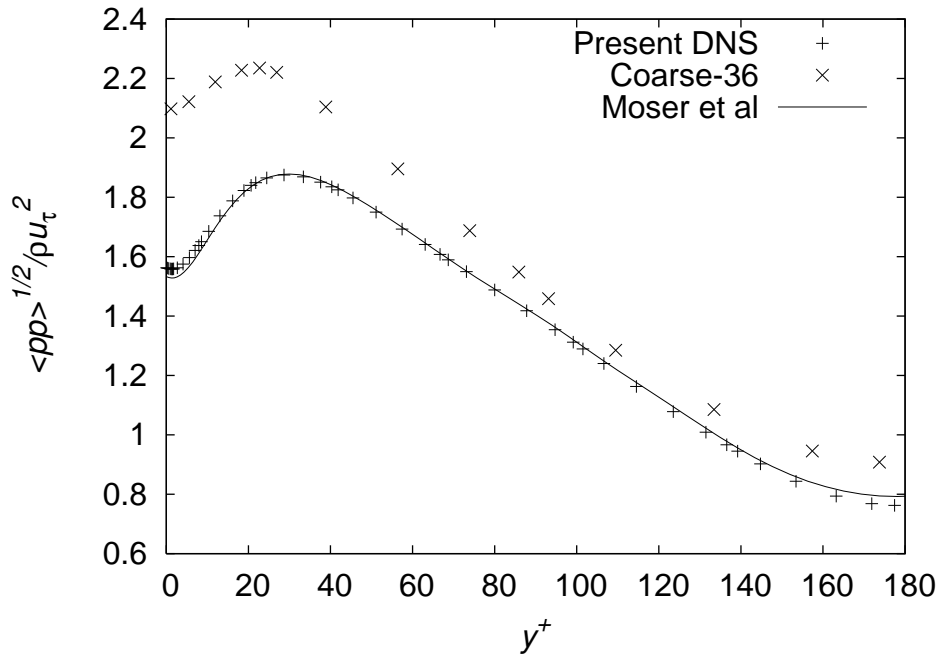}
     \caption{$\mathrm{Re}_\tau = 180$: Variation of the
            root-mean-square pressure fluctuations 
            across half the channel, compared with the reference solution of 
            Moser \etal~\protect{\cite{RDMoser_JKim_NNMansour_1999a}}.}
   \label{fig:prms}
\end{figure}

The actual computed Reynolds number is $\mathrm{Re}_*=178.83$, 
i.e. within 0.7\% of the prescribed value and well within what can be
expected. Moser \etal~%
\cite{RDMoser_JKim_NNMansour_1999a} reported $\mathrm{Re}_*=178.13$. 
The results presented in Figs.~\ref{fig:U_log}--\ref{fig:prms}
compare in all aspects very well with the benchmark data, thus
establishing solid confidence in the numerical method. The slight
deviations reported herein is well within what should be expected, and
even closer correspondence could have been obtained by simply
collecting statistics for a longer period of time. This was, however,
not considered to be necessary. 

As background for the VMS LES results presented below, we also
include results from a simulation on the grid ``Coarse-36'' (see
Table~\ref{table:vmsgrids} for grid properties). This simulation contains
no turbulence modelling, but 2\%\ polynomial
filtering~\cite{PFFischer_JSMullen_2001a} is employed. Except for the
pressure correlations in Fig.~\ref{fig:prms}, the results are so good
that modelling is not expected to improve them. This shows that the
spectral element method gives high accuracy even for relatively
coarse grids, but it also indicates that plane channel flow is not the
most challenging test case. The availability of quality reference data
makes it attractive as a starting case, we must however keep in mind
that the grids for the model tests have to be sufficiently coarse and
not turn into a ``quasi-DNS'' e.g.\ near the walls.

%\clearpage

\subsection{VMS LES results}

Lots of combinations of the scale partitioning parameter and the
Smagorinsky forms were tested for $\mathrm{Re}_\tau=180$, and the best
choice was used for additional simulations at $\mathrm{Re}_\tau=550$
and $\mathrm{Re}_\tau=950$.

The spectral element grid for $\mathrm{Re}_\tau=180$ was chosen as the
``Coarse-24'' grid described in Table~\ref{table:vmsgrids}.  The
element interfaces in the wall-normal direction were given by a coarse  
Gauss-Lobatto-Chebyshev grid, as recommended
in~\cite{GEKarniadakis_SJSherwin_1999a}. The scale partitioning
cut-off mode was kept constant for all elements, even though the
element size varied in the wall-normal direction.

In order to get a real test of the modelling, the grid had to be much
coarser than what would give reasonably good results without a
model. Spectral element grids for the higher Reynolds numbers were
constructed such that the first element interface in the wall-normal
direction is placed at approximately the same value of $y^+$ for all
the cases, see the illustrations in Fig.~\ref{fig:ygrids}. To reduce
the number of parameters, the polynomial degree was fixed for all the
VMS LES runs; only the number of elements was changed.

\subsubsection{Simulations at $\mathrm{Re}_\tau=180$}

The grid parameters for this case are given in the column ``Coarse-24''
in Table~\ref{table:vmsgrids}.

Without a model, both over-integration and polynomial filtering (2\%)
was necessary to keep the simulation stable at this resolution. With
the VMS model term, either method was sufficient. It was found that
polynomial filtering did reduce rather than improve the quality of the
results. To obtain the presented VMS results we therefore employed
only over-integration in the simulations.

Beside using the different forms of the Smagorinsky
term~\eqref{eq:small-small-smagorinsky}--\eqref{eq:full-small-smagorinsky},
the scale partitioning was varied in the simulations. With a local
grid of 7 grid points in each direction on each element, we have used
$\overline{N}=4$ and $\overline{N}=5$ for the large-scale extraction
described in Section~\ref{sec:nodal-modal}. These values correspond to
57\%\ and 71\%\ of the one-dimensional spectrum, respectively. In
three dimensions, the resulting large-scale spaces consist of 19\%\
and 35\%\ of the modes, respectively.

\begin{figure}
     \includegraphics[scale=1,angle=0]{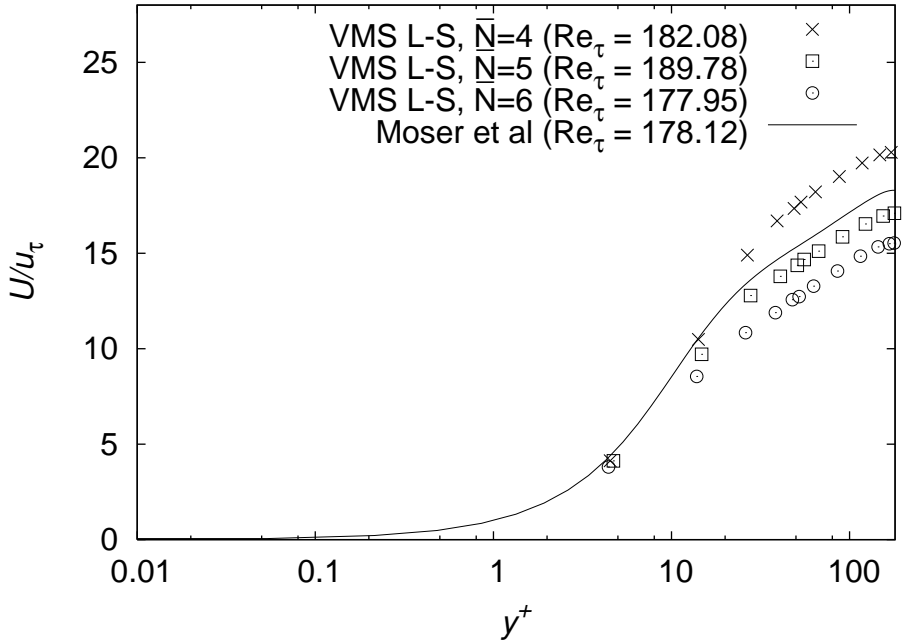}
     \caption{$\mathrm{Re}_\tau = 180$: Variation of the mean velocity
       across half the channel in viscous units, compared with the
       reference solution of Moser et
       al.~\protect{\cite{RDMoser_JKim_NNMansour_1999a}}.} 
   \label{fig:180lf_U_log}
\end{figure}

\begin{figure}
    \includegraphics[scale=1,angle=0]{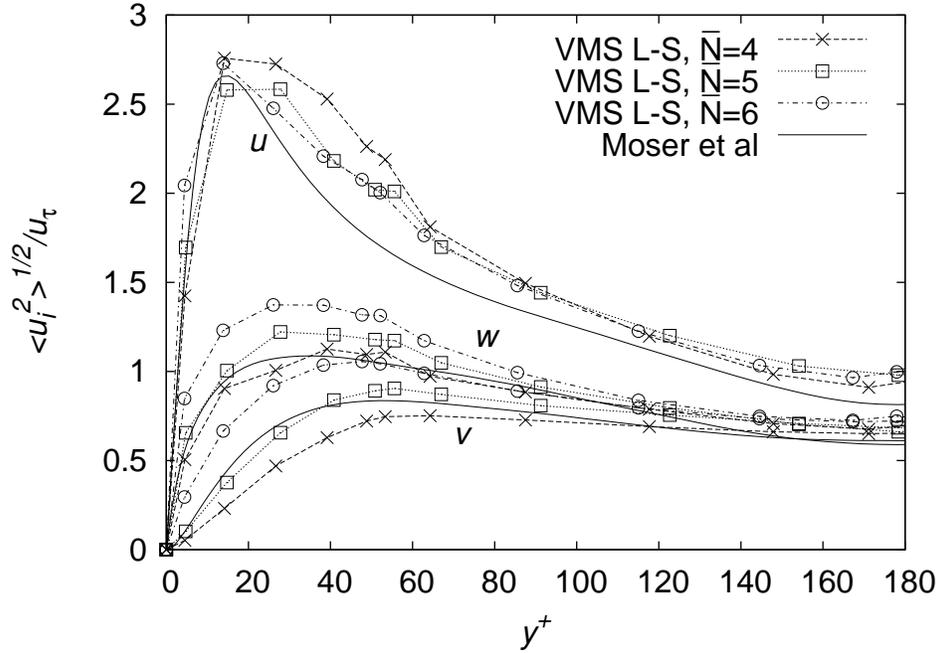} 
    \caption{$\mathrm{Re}_\tau = 180$: Variation of streamwise ($u'$),
            spanwise ($w'$), and wall-normal ($v'$)  
            root-mean-square velocity fluctuations 
            across half the channel, compared with the reference solution of 
            Moser \etal~\protect{\cite{RDMoser_JKim_NNMansour_1999a}}.}
    \label{fig:180lf_urms}
\end{figure}

Varying the scale partitioning had a strong influence on the results,
and $\overline{N}=5$ was found to be the best choice, as seen from
Figs.~\ref{fig:180lf_U_log} and~\ref{fig:180lf_urms}. The rest of the
results shown here are obtained with $\overline{N}=5$.

\begin{figure}
     \includegraphics[scale=1,angle=0]{fig11.eps}
     \caption{$\mathrm{Re}_\tau = 180$: Variation of the mean velocity
       across half the channel in viscous units, compared with the
       reference solution of 
       Moser \etal~\protect{\cite{RDMoser_JKim_NNMansour_1999a}}.}
   \label{fig:180_U_log}
\end{figure}

\begin{figure}
     \includegraphics[scale=1,angle=0]{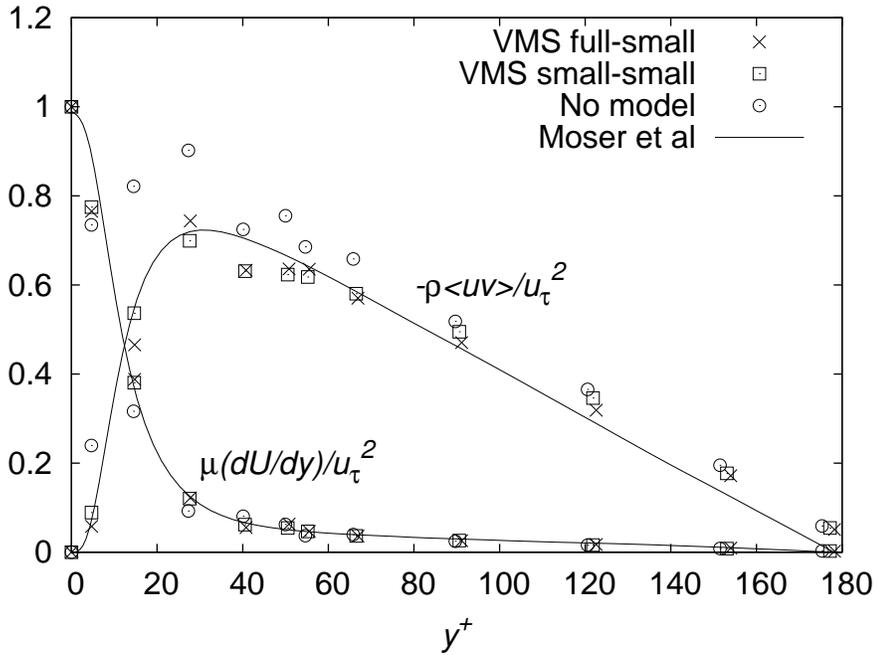}
     \caption{$\mathrm{Re}_\tau = 180$: Variation of mean viscous
            shear and the turbulent shear stress  
            across half the channel, compared with the reference solution of 
            Moser \etal~\protect{\cite{RDMoser_JKim_NNMansour_1999a}}.}
    \label{fig:180_shear_stress} 
\end{figure}

\begin{figure}
    \includegraphics[scale=1,angle=0]{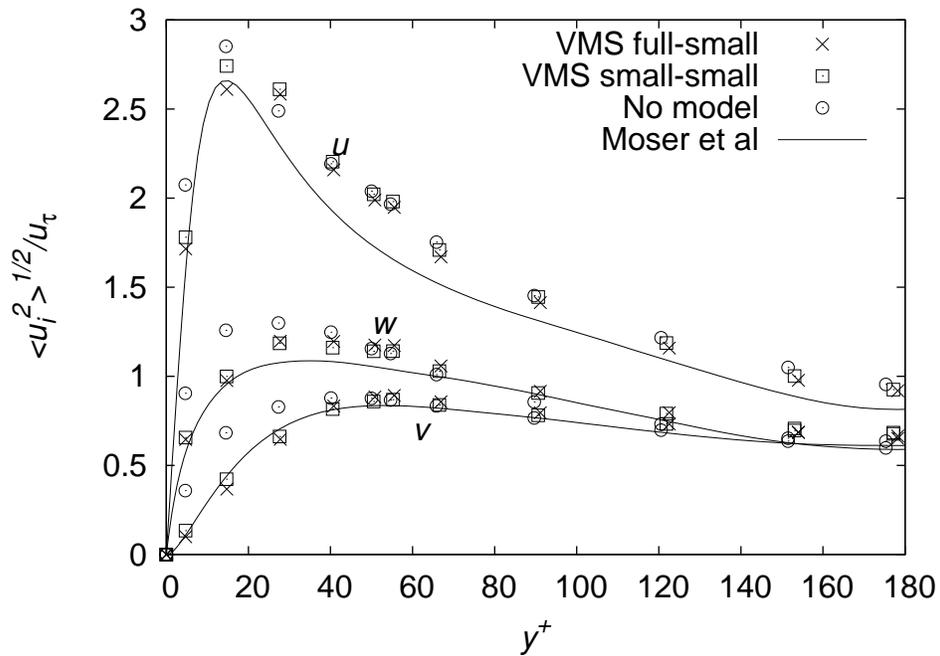} 
    \caption{$\mathrm{Re}_\tau = 180$: Variation of streamwise ($u'$),
            spanwise ($w'$), and wall-normal ($v'$)  
            root-mean-square velocity fluctuations 
            across half the channel, compared with the reference solution of 
            Moser \etal~\protect{\cite{RDMoser_JKim_NNMansour_1999a}}.}
    \label{fig:180_urms}
\end{figure}

\begin{figure}
     \includegraphics[scale=1,angle=0]{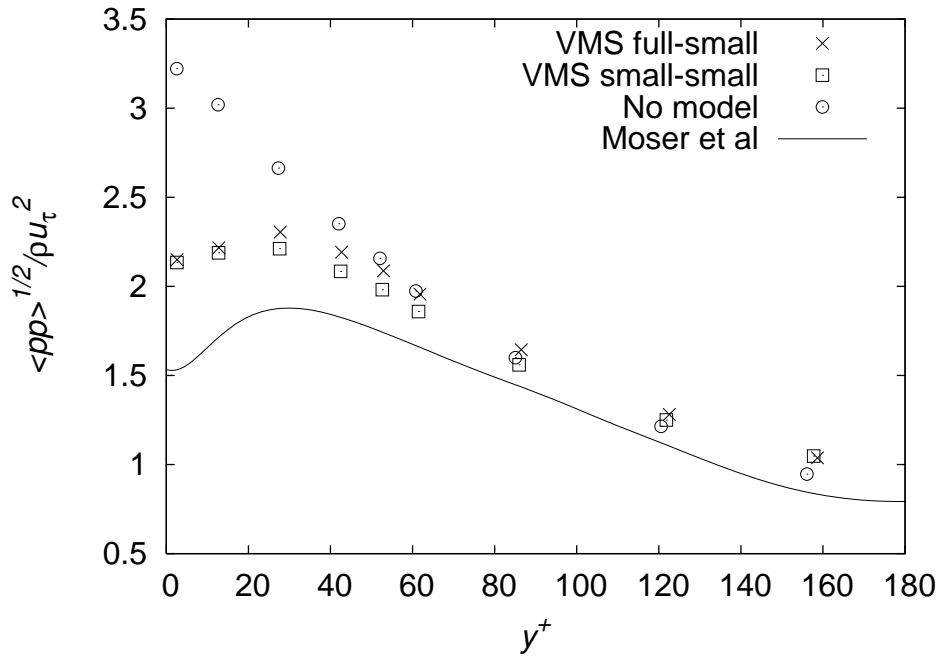}
     \caption{$\mathrm{Re}_\tau = 180$: Variation of the
            root-mean-square pressure fluctuations 
            across half the channel, compared with the reference solution of 
            Moser \etal~\protect{\cite{RDMoser_JKim_NNMansour_1999a}}.}
   \label{fig:180_prms}
\end{figure}

The different forms of the Smagorinsky term gave very similar results
for $\mathrm{Re}_\tau=180$. The results are presented in
Figs.~\ref{fig:180_U_log}--\ref{fig:180_prms}. The results from
``large-small'' form~\eqref{eq:large-small-smagorinsky} were almost
indistinguishable from the
``full-small''~\eqref{eq:full-small-smagorinsky} results, and are not
included in the figures.

As shown in Section~\ref{dns-results}, simulations on the
``Coarse-36''-grid gave good results without modelling for this
case. Results from simulations without modelling on an intermediate
grid with $30^3$ grid points were comparable to the VMS results from
the $24^3$-grid shown here, but at a 40\%\ higher computational cost.

\subsubsection{Simulations at $\mathrm{Re}_\tau=550$}

\begin{figure}
     \includegraphics[scale=1,angle=0]{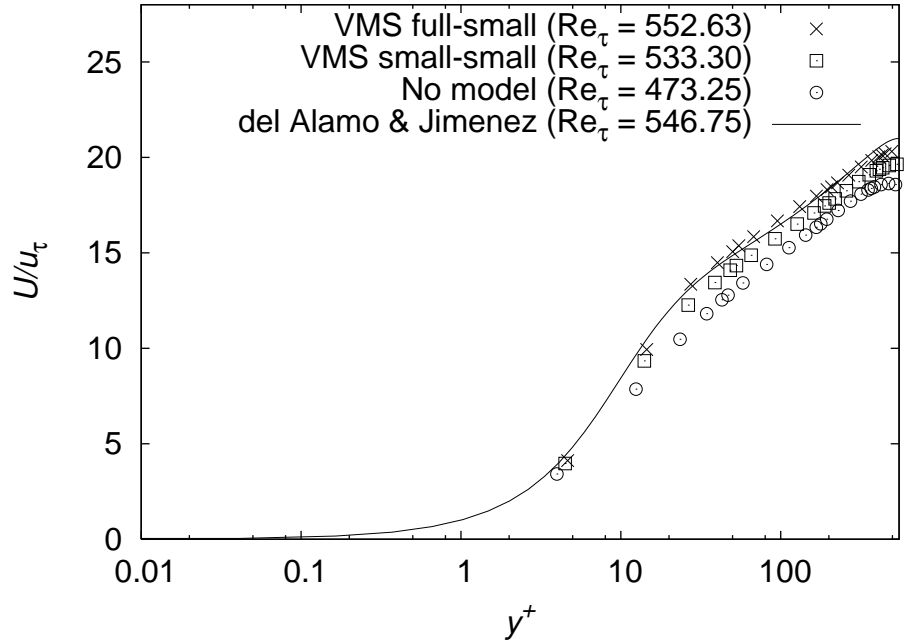}
     \caption{$\mathrm{Re}_\tau = 550$: Variation of the mean velocity
    across half the channel in viscous units, compared with the
    reference solution of del \'{A}lamo and
    Jim\'{e}nez~\protect{\cite{JCdelAlamo_JJimenez_2003}}.} 
   \label{fig:550_U_log}
\end{figure}

\begin{figure}
     \includegraphics[scale=1,angle=0]{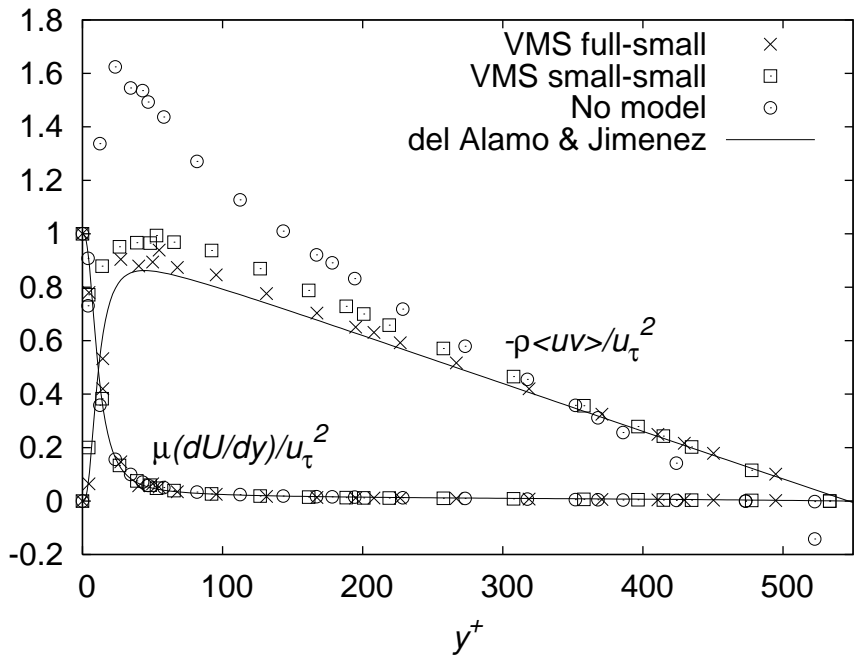}
     \caption{$\mathrm{Re}_\tau = 550$: Variation of mean viscous
            shear and the turbulent shear stress across half the
            channel, compared with the reference solution of  
            del \'{A}lamo and Jim\'{e}nez~\protect{\cite{JCdelAlamo_JJimenez_2003}}.}
    \label{fig:550_shear_stress} 
\end{figure}

\begin{figure}
    \includegraphics[scale=1,angle=0]{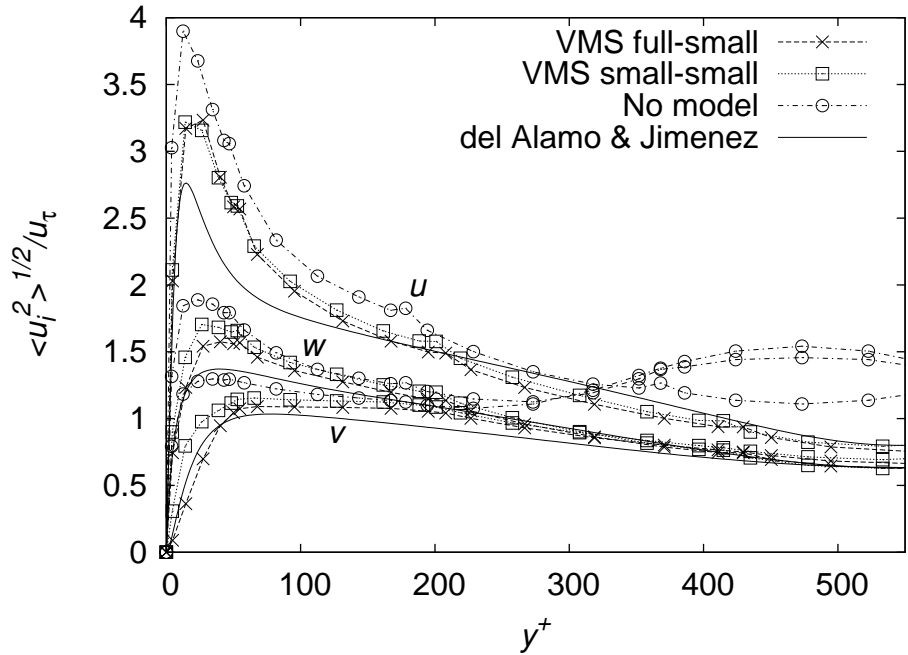} 
    \caption{$\mathrm{Re}_\tau = 550$: Variation of streamwise ($u'$),
            spanwise ($w'$), and wall-normal ($v'$)  
            root-mean-square velocity fluctuations 
            across half the channel, compared with the reference solution of 
            del \'{A}lamo and
            Jim\'{e}nez~\protect{\cite{JCdelAlamo_JJimenez_2003}}.}
    \label{fig:550_urms}
\end{figure}

The grid parameters for this case are given in the column ``Coarse-42''
in Table~\ref{table:vmsgrids}.

The scale partitioning parameter of $\overline{N}=5$, which was found
to be the best choice for $\mathrm{Re}_\tau=180$, was also used for
this case. Again, the ``full-small'' and ``large-small'' Smagorinsky
forms produced very similar results, so the latter are not shown. The
results are presented in Figs.~\ref{fig:550_U_log}--\ref{fig:550_urms}.

\subsubsection{Simulations at $\mathrm{Re}_\tau=950$}

\begin{figure}
     \includegraphics[scale=1,angle=0]{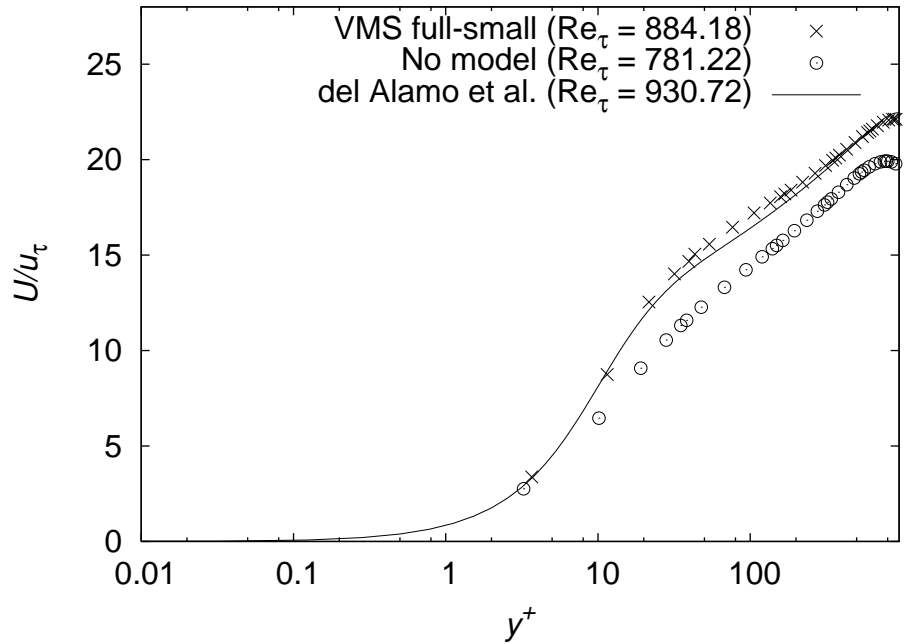}
     \caption{$\mathrm{Re}_\tau = 950$: Variation of the mean velocity
    across half the channel in viscous units, compared with the
    reference solution of del \'{A}lamo et
    al.~\protect{\cite{JCdelAlamo_etal_2004a}}.}  
   \label{fig:950_U_log}
\end{figure}

\begin{figure}
     \includegraphics[scale=1,angle=0]{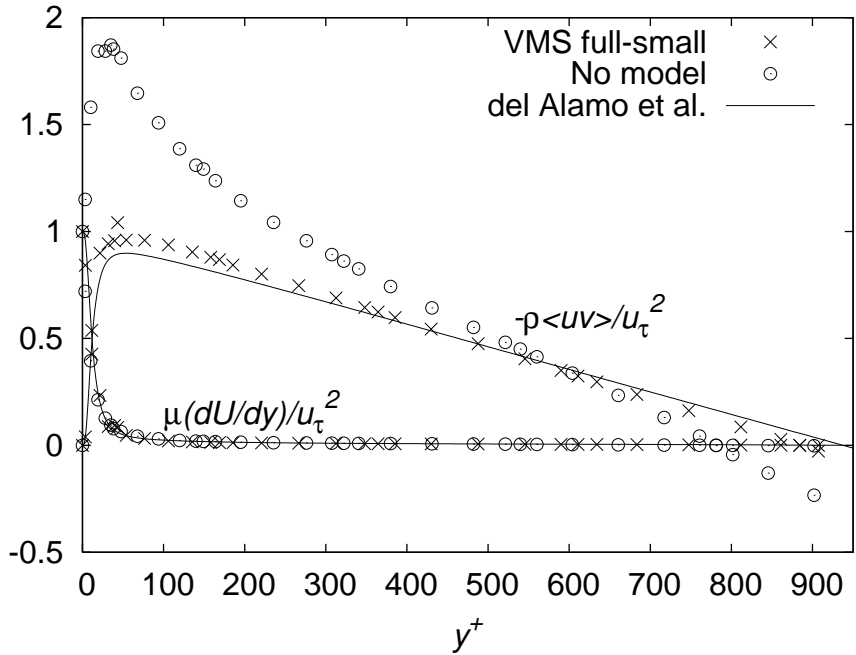}
     \caption{$\mathrm{Re}_\tau = 950$: Variation of mean viscous
            shear and the turbulent shear stress across half the
            channel, compared with the reference solution of  
            del \'{A}lamo \etal~\protect{\cite{JCdelAlamo_etal_2004a}}.}
    \label{fig:950_shear_stress} 
\end{figure}

\begin{figure}
    \includegraphics[scale=1,angle=0]{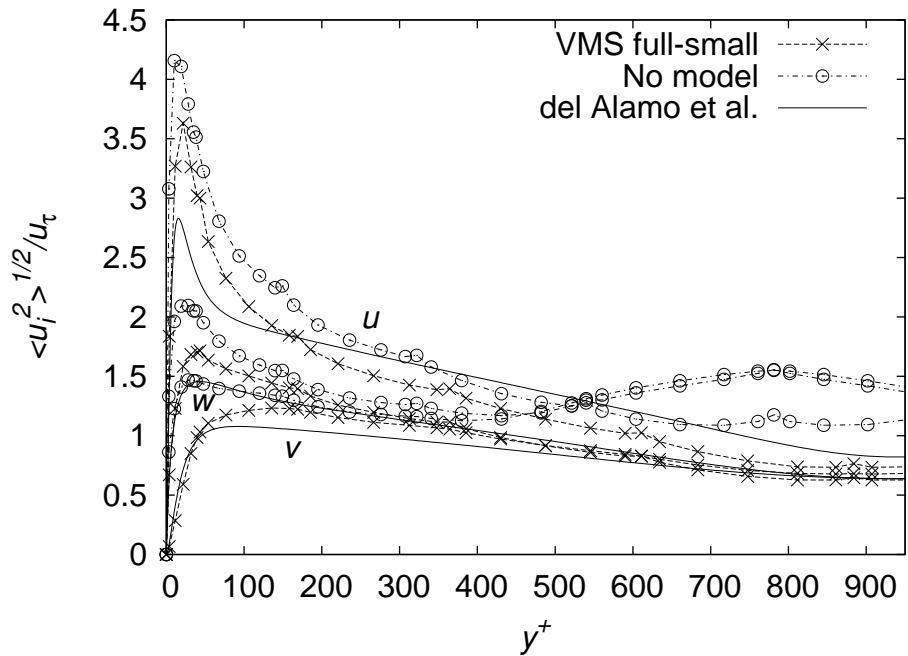} 
    \caption{$\mathrm{Re}_\tau = 950$: Variation of streamwise ($u'$),
            spanwise ($w'$), and wall-normal ($v'$)  
            root-mean-square velocity fluctuations 
            across half the channel, compared with the reference solution of 
            del \'{A}lamo \etal~\protect{\cite{JCdelAlamo_etal_2004a}}.}
    \label{fig:950_urms}
\end{figure}

The grid parameters for this case are given in the column ``Coarse-60''
in Table~\ref{table:vmsgrids}.

In this case we have only run the ``full-small''  Smagorinsky form,
and the scale partitioning parameter is still $\overline{N}=5$.
The reference simulation is described in~\cite{JCdelAlamo_etal_2004a},
and the reference data are downloaded from the site given
in~\cite{JCdelAlamo_JJimenez_2003}. 
Our results are presented in Figs.~\ref{fig:950_U_log}--\ref{fig:950_urms}.

\subsection{Comments on the results}

The VMS LES results show clear improvement from the results without a
model, in particular for the higher Reynolds numbers. The plane
channel flow at $\mathrm{Re}_\tau=180$ does not seem to provide
sufficient challenges for the testing of turbulence models, as it is
too easy to resolve the main features without any modelling at
all. The VMS LES results are not compared with alternative turbulence
models, as the intentions of this study was mainly to lay the
foundations for the incorporation of VMS LES in a spectral element
method. Therefore only the simplest Smagorinsky eddy viscosity was
used in the model terms in the small-scale equations.

% -----------------------------------------------------------------------------
\section{Conclusions}
\label{sec:conclu}

The Variational Multiscale Large Eddy Simulation method has been
implemented within the framework of a spectral element method. The
presented scale partitioning method was shown to produce a
gradual introduction of the small-scale model terms. This is
intuitively favourable to a sharp cut-off at a given point in the
spectral space. The computational overhead for the method was
20--40\%\ for the applications considered here. This must be
considered to be reasonably low, as even small increases in the
spatial resolution of the spectral element method are more
computationally demanding. Good results have been obtained for plane
channel flows up to $\mathrm{Re}_\tau=950$, even for grid densities as
low as 0.06\%\ of the reference simulation grid density, and using the
simplest possible small-scale dissipation model.

% -----------------------------------------------------------------------------

\begin{ack}
This work was in part performed under the WALLTURB project. 
WALLTURB (A European synergy for the assessment of wall turbulence) is funded 
by the CEC under the 6th framework program (CONTRACT No: AST4-CT-2005-516008).

This work received support from the Research Council of
Norway (Programme for Supercomputing) through a grant of
computing time.

We are grateful to Professor Lars Davidson for providing suitable
initial data for the plane channel flow simulations.

\end{ack}

%\bibliographystyle{elsart-num}
%\bibliography{references}

\end{document}